\documentstyle[prl,aps,preprint,tighten,psfig]{revtex}
\def\sech{\hbox{sech}}
\begin{document}
\draft
\title{Coherent Charge Transport in Superconductor -- Normal Metal Proximity
Structures}
\author{A.A.Golubov$^{a,b}$, F.Wilhelm$^c$ and A.D.Zaikin$^{c,d}$}
\address{$^a$ Institute of Thin Film and Ion Technology,\\Research Centre J\"ulich
(KFA), D-52425 J\"ulich, Germany,\\ $^b$ Institute of Solid State Physics,
142432 Chernogolovka, Russia\\ $^c$ Institut f\"ur Theoretische
Festk\"orperphysik, Universit\"at\\Karlsruhe, 76128 Karlsruhe, FRG \\ $^d$
I.E.Tamm Department of Theoretical Physics,\\P.N.Lebedev Physics Institute,
Leninskii prospect 53, 117924\\Moscow, Russia }
\maketitle

\begin{abstract}
We develop a detailed analysis of electron transport in normal diffusive
conductors in the presence of proximity induced superconducting correlation.
We calculated the linear conductance of the system and the profile of the
electric field. A rich structure of temperature dependences was found
and explained. Our results are consistent with recent experimental findings.
\end{abstract}

%\pacs{PACS numbers: }

%\twocolumn[
%\hsize\textwidth\columnwidth\hsize\csname @twocolumnfalse\endcsname

%]
\newpage

\narrowtext
Recent progress in nanolithographic technology revived the interest to both
experimental and theoretical investigation of electron transport in various
mesoscopic proximity systems consisting of superconducting and normal
metallic layers. In such systems the Cooper pair wave function of a
superconductor penetrates into a normal metal at a distance which increases
with decreasing temperature \cite{dG}. At sufficiently low temperatures this
distance becomes large and the whole normal metal may acquire
superconducting properties. Although this phenomenon has been
already understood more than thirty years ago and intensively investigated
during past decades, recently novel physical features of metallic proximity
systems have been discovered \cite{Kas,Petr,Pot,Cour} and studied
theoretically (see \cite{Zai90,Volk92,VZK,Been92,Been94,HN,Naz,Zaik,Spivak,WSZ}
and further references therein).

In this paper we study the influence of the proximity effect on transport
properties of a diffusive conductor in the limit of relatively low
temperatures and voltages. We will assume that this conductor is brought in
a direct contact to a superconducting reservoir which serves as an effective
injector of Cooper pairs into a normal metal. It is reasonable to assume
that in this case the system conductance can only increase in comparison to
its normal state value. As the penetration length of Cooper pairs into a
normal metal becomes larger with decreasing temperature $T$ one might expect
the system conductance to increase with $1/T$ monotoneously. In what follows
we shall present a rigorous calculation of a temperature dependent linear
conductance of the the system and demonstrate that its actual temperature
dependence turns out to be more complicated. In particular we will show that
if the system contains no tunnel barriers there are two different physical
regimes which determine the system conductance in different temperature
intervals. At relatively high temperatures the superconducting correlation 
of electrons in the normal metal survives at the typical distance of order
$\xi_N \sim \sqrt{{\cal D}/T}$, where ${\cal D}=v_Fl_{imp}/3$ is the diffusion 
coefficient.
As $T$ is lowered proximity induced superconductivity expands into the
normal metal and, consequently, the ``normally conducting'' part of the system 
effectively shrinks in size. This effect results in increasing of the 
conductance of a normal metal. At sufficiently low temperature the length
$\xi_n$ becomes of order of the size of the normal layer, i.e. the 
superconducting correlation survives everywhere in the normal metal and the
effective gap in the quasiparticle spectrum develops there \cite{GK}. Physically
this means that at sufficiently low energies and temperatures there exist
no uncorrelated electrons in the normal metal with a superconducting
reservoir attached to it. The characterisic energy which plays the role
of such a gap is $\varepsilon_g\sim \min (\Delta ,D/L^2)$. 
At energies lower than this value electron transport is only due to
such correlated electrons (or Cooper pairs injected into the N-metall)
which ``density of states'' increases with increasing $\varepsilon$ due to the
gap effect. As the temperature is increased from zero higher and higher values
of $\varepsilon$ contribute to the current and the system conductance
increases with $T$ at very low temperatures. Eventually with increasing
$T$ a crossover to a high temperature behavior of the system conductance
takes place. Note that similar behavior 
of the normal metal conductance in the presence of proximity-induced 
superconductivity has been also found in a recent
numerical analysis by Nazarov and Stoof \cite{NazSt}. 
It is also interesting to point out that at $T=0$ the system
conductance coincides with that of a normal metal with no proximity effects
\cite{AVZ}. 

We will also demonstrate that in the presence of tunnel barriers
the temperature dependence of the conductance is essentially different.
Whereas the contribution of the diffusive region is nonmonotoneus in T, the
conductance of the interfaces decreases monotonously with temperature. At
the same time the proximity-induced superconducting gap developes in N 
\cite{GK}. As a result, with lowering the barrier transparency the crossover
takes place to the effective conductance decreasing monotonously with T,
characteristic for two serial NIS' tunnel junctions (S' is now the diffusive
normal conductor with proximity-induced gap). We note that both types of
behavior, namely nonmonotonously enhanced and monotonously decreasing with T
conductance have been observed in the experiment \cite{Petr}.
 
Let us consider a a quasi-one-dimensional normal conductor of a length $2L$
with a superconducting strip of a thickness $2d_s$ attached to a normal metal
on the top of it and two normal reservoirs attached to its edges (see fig.1). The length $L$ is assumed to be much larger than the elastic mean free
path $l_{imp}$ but much shorter than the inelastic one. This
geometrical realization has a direct relation to that investigated in the
experiments \cite{Petr,Cour}. The two big normal reservoirs N' are assumed to be in
thermodynamic equilibrium at the potentials $V$ and $0$ respectively. In contrast to the case
of ballistic constriction \cite{BTK} the potential drop within the system is
distributed between the interfaces and the conductor itself. The general
approach to calculate the conductance of such kind of structures was
developed in \cite{Zai90,Volk92,VZK}. In what follows we shall apply this
method to analyse the temperature dependence of the NS proximity structure
of fig.1. 

The electron transport through the metallic system can be described by the
equations for a matrix of quasiclassical Green functions$%
\stackrel{\vee }{G\text{ }}$in the contact \cite{Eliash,LO}: 
\begin{equation}
\label{G}\stackrel{\vee }{G\text{ }}=\left( 
\begin{array}{cc}
\stackrel{\wedge }{G}^R & \stackrel{\wedge }{G}^K \\ 0 & \stackrel{\wedge }{G%
}^A
\end{array}
\right) 
\end{equation}
where $\stackrel{\wedge }{G^A}$, $\stackrel{\wedge }{G^R}$ and $\stackrel{%
\wedge }{G^K}$ are respectively the impurity-averaged advanced, retarded and 
Keldysh Green functions. These functions are in turn matrices in the Nambu space:
$$
\stackrel{\wedge }{G^R}=%
\stackrel{\wedge }{\sigma _z}g^R+\stackrel{\wedge }{i\sigma _y}f^R,
\stackrel{\wedge }{G^A}=-(\stackrel{\wedge }{G^R})^{*}\hbox{and}\stackrel{%
\wedge }{G^K}=\stackrel{\wedge }{G^R}\stackrel{\wedge }{f}-\stackrel{\wedge 
}{f}\stackrel{\wedge }{G^A}
$$. Here the distribution function $%
\stackrel{\wedge }{f}=f_l+\stackrel{\wedge }{\sigma _z}\stackrel{\wedge }{f_t%
}$, where $f_l=\tanh (\varepsilon /2T)$ and $\stackrel{\wedge }{f_t}$
describes deviation from nonequilibrium. Taking advantage of the
normalization condition for the normal and the anomalous Green functions 
$(g^R)^2-|f^R|^2=1$  it is convenient to parametrize 
$g^R=\cosh \theta ,$ $f^R=\sinh \theta $, where $\theta \equiv \theta
_1+i\theta _2$ is a complex function. Deep in the bulk
superconductor it is equal to $\theta _s=1/2\ln \left[ (\Delta +\varepsilon
)/(\Delta -\varepsilon )\right]-i\pi/2 $  for $\varepsilon <\Delta $ and $\theta
_s=(1/2)\ln \left[ (\varepsilon +\Delta )/(\varepsilon -\Delta )\right] /2$
for $\varepsilon >\Delta $ (here and below we omit the indices R(A)).

The electrical current $I$ and the electrostatic potential $\phi$ are expressed 
through $\stackrel{\vee }{G\text{ }}$ as
\begin{equation}
\label{curr-d}I=\frac{\nu {\cal D}S}2\int_{-\infty }^\infty d\varepsilon 
\text{ }Sp\left[ \stackrel{\wedge }{\sigma _z}\stackrel{\vee }{G\text{ }}%
\partial _x\stackrel{\vee }{G\text{ }}\right] ^K,
\end{equation}
\begin{equation}
\phi(x)=\int_0^\infty d\varepsilon\,\hbox{Tr}\hat{g}^K(x,\varepsilon)
=\int_0^\infty d\varepsilon\,f_t(x,\varepsilon)\nu_\varepsilon (x),
\end{equation}
where $\nu $ is the the density of 
states, $\nu_\varepsilon(x)=\Re(g^R_\varepsilon(x))$ and $S$ is the crossection area of $N$ conductor. 

Being expressed in terms of the function $\theta (\varepsilon ,x)$ the 
equations \cite{Eliash,LO} for the Green functions and the distribution 
function for the N-metall take a particularly simple form 
\begin{equation}
\label{eq1}{\cal D}\partial _x^2\theta +2i\varepsilon \sinh \theta =0
\end{equation}

\begin{equation}
\label{eq2}\partial _x\left[ {\cal D}(\cosh ^2\theta _1)\partial
_xf_t\right] =0,
\end{equation}
$x$ is the coordinate along the N-conductor. Here we neglected the processes of inelastic relaxation and put the pair
potential in the normal metal equal to zero $\Delta _N=0$ assuming
the absence of electron-electron interaction in this metal.

Before we come to a detailed solution of the problem let us point out
that the conclusion about the anomalous behavior of the system conductance
can be reached already from the form of eq. (\ref{eq2}). Indeed it is
quite clear from (\ref{eq2}) that the effective diffusion coefficient
increases in the N-regions with proximity-induced superconductivity and, therefore,
the electric field is partially expelled from these regions. This 
energy dependent field modulation is controlled by the solution for $\theta
(\varepsilon ,x)$ and is directly related to the physical origin of the 
anomalous temperature dependence of the system conductance discussed below. 

The equations (\ref{eq1}) and (\ref{eq2}) should be supplemented by the boundary
conditions at the interfaces of the normal metall N. Assuming that the anomalous
Green function of big normal reservoirs N' is equal to zero from \cite{KL,Volk92}
we obtain 
\begin{equation}
\label{BC}
\begin{array}{c}
\gamma _B\partial _x\theta =\sinh \theta  \\ 
\gamma _B\cosh \theta _1\partial _xf_t=\cosh \theta _1(f_t-f_t(x=0,2L)),
\end{array}
\end{equation}
where $\gamma _B=R_b/\rho _N$ is the interface resistance parameter,  $R_b$
is the resistance of the interface between the N-conductor and the 
N'-reservoirs, $\rho _N$ is the resistivity of the N-metal. In general we
should also fix the boundary condition at the interface between the N-metal and
the superconductor. For the case of a perfect transparency of this interface
(which is only considered here) and for typical thickness of the normal
layer $w_N \sim \sqrt{S} \\ \xi_N$ Cooper pairs easily penetrate into it due to 
the proximity effect and the Green functions
of the N-metal at relatively low energies for $d\le x\le d+2d_s$ are equal to 
those of a bulk superconductor $\theta =\theta _s$. This intuitively obvious 
result can be proven rigorously (see e.g. \cite{WSZ} and references therein). Thus the region
of a normal metal situated directly under the superconductor for our
purposes can be also considered as a piece of a superconductor S' and the 
solution of (\ref{eq1}), (\ref{eq2}) needs to be found only for $0<x<d$
(without loss of generality we will stick to a symmetric configuration). 

Proceeding along the same lines as it has been done in ref. \cite{VZK} 
we arrive at the final expression for the current
\begin{equation}
\label{current}
I=\frac 1{2R}\int_0^\infty d\varepsilon \left[ \tanh ((\varepsilon
+eV)/2T)-\tanh ((\varepsilon +eV)/2T)\right] D(\varepsilon ), 
\end{equation}
where $D(\varepsilon )$ defines the effective transparency of the system 
\cite{VZK}
$$
D(\varepsilon )=\frac{L+\gamma _B}{\frac{\gamma _B}{\cosh \theta
_2(x=0,\varepsilon )\cos \theta _1(x=0,\varepsilon )}
+\int_0^Ldx\,\sech^2\theta _1(x,\varepsilon )}. 
$$
Let us consider the case of a sufficiently long normal conductor 
$d \gg {\cal D}/\Delta$. Then at low temperatures $T \ll \Delta$
the interesting energy interval is restricted to $\varepsilon\ll\Delta$.
For such values of $\varepsilon$ the contribution of the $S^\prime$-part 
of the normal conductor shows no structure and can be easily taken into account
with the aid of obvious relations
\begin{equation}
\int_0^Ldx\,\sech^2\theta_1(x,\varepsilon)=\int_0^{d}dx\;\sech^2\theta_1(x,\varepsilon)+d_s\sech^2\theta_{s,1}
\end{equation}
and $\sech^2\theta_{s,1}=\left(1-{\varepsilon^2\over\Delta^2}\right)$. 
Due to this reason we will discuss only the properties of the $N$-part ($0<x<d$).
For the sake of completeness we will also demonstrate the effect of finite 
$d_s$ in the end of our calculation.

For the differential conductance of the $N$-part $0\le x\le d$ 
normalized to its normal (``non-proximity'') value in the zero bias limit 
eq. (\ref{current}) yields
\begin{equation}
\label{conduct}
\bar{\sigma}_N=\left({RdI\over dV}\right)_{V=0}={1\over2T}\int_0^\infty
d\varepsilon{D(\varepsilon)\sech^2(\varepsilon/2T)}.
\end{equation}
Analogously the normalized zero-bias electrostatic potential distribution
reads
\begin{equation}
\label{potential}
\phi_0(x)=\lim_{V\rightarrow0}{\phi(x)\over V}
={1\over 2Td}\int_0^\infty d\varepsilon\,
D(\varepsilon)\nu_\varepsilon(x)\sech^2(\varepsilon/2T)\int_x^{d}dx^\prime\,\sech^2(\theta_1(x^\prime))
\end{equation}

The analysis of the problem can be significantly simplified in the case of 
perfectly transparent interfaces ($\gamma_B=0$). In this case the boundary
conditions are 
\begin{eqnarray}
\label{bc1}
\theta(0)&=&0\\
\theta(d)&=&\theta_S
\end{eqnarray}
for the contact to the normal and the superconducting reservoir
respectively. The effective transparency of the N-part then reads
\begin{equation}
D(\varepsilon)=\left({1\over
d}\int_0^{d} dx\;\sech^2(\theta_1(x))\right)^{-1}.
\end{equation}
As it was already pointed out for relatively long normal conductors and 
at low $T$ only the energies $\varepsilon\ll\Delta$ give an 
important contribution to the conductance. In this case the typical energy
scale is defined by the Thouless energy $\epsilon_d={\cal{D}/d^2} \ll \Delta$.
For these energies we can set $\theta_S=i{\pi/2}$.
Let us first put $T=0$. Then the thermal distribution factor
$\beta\,\sech^2(\beta\varepsilon)$ reduces to a delta function and we have
\begin{equation}
\bar{\sigma}_N(T=0)=D(0),
\end{equation}
\begin{equation}
\phi_0(x)=(D(0)\nu_0(x)/d)\int_x^{d}dx^\prime\sech^2(\theta_1(x^\prime)),
\end{equation}
i.e. we only need the solution of (\ref{eq1}) with boundary conditions
(\ref{bc1}) at $\varepsilon=0$, which is
$\theta=i{\pi\over2}{\bar{x}}$ and therefore
$\nu_0(x)=\cos(\bar{x}\pi/2)$, where $\bar{x}=x/d$. This
means 
\begin{equation}
\bar{\sigma}_N(T=0)=1.
\end{equation} 
This result coincides with that obtained first by Artemenko, Volkov and
Zaitsev \cite{AVZ} and demonstrates that -- in contrast 
to what one might expect intuitively -- at $T=0$ and very low voltages 
the diffusive conductor does not ``feel'' the proximity-induced
superconductivity and the conductance is exactly equal to its 
normal state value. The profile of the electric field penetrating
into the normal metal at $T=0$ also can be found easily. We have
\begin{equation}
E_0(x):={\partial_{\bar{x}}}\phi_0(x)
={\pi\over2}(\bar{x}-1)\sin(\bar{x}\pi/2)-\cos(\bar{x}\pi/2),
\end{equation}
i.e. in the low temperature limit the electrical field distribution 
in the structure turns out to be essentially nonmonotoneous. 
In the case $T\ll\varepsilon_d$ we can calculate $\theta$ perturbatively.
From $\cal{D}\partial_x^2\theta=-2i\varepsilon\sinh\theta_0(x)$ and
(\ref{bc1}) we get 
$$
\theta={8\over\pi^2}{\varepsilon\over
\varepsilon_d}[\bar{x}-\sin(\bar{x}\pi/2)]+i{\pi\over2}\bar{x}.$$
Keeping only leading order terms in
$\epsilon\over\epsilon_d$, we get 
\begin{equation}
\bar{\sigma}_N=1+A{T^2\over\epsilon_d^2},
\end{equation}
where
$A={128\over\pi^4}\left({5\over6}-{8\over\pi^2}\right)\zeta(2)\approx0.049$
is a universal constant. This means, that for low temperatures
$\bar{\sigma}_N(T)$ grows quadratically on the scale of
$\varepsilon_d$ and approaches the crossover towards the high temperature
regime discussed below.

We can also calculate the electrical field distribution in this
approximation, however, the result is rather cumbersome and shows no
structure which could not be seen from the numerical
results presented below.

In this limit $T\gg\epsilon_d$ (where we still have $T\ll\Delta$), the 
contribution of the low energy components to the
thermally weighted integrals for $\bar{\sigma}_N(T)$ and $\phi_0(T)$
is neglible and we only have to take into account the solutions of (\ref{eq1}) 
for energies $\epsilon\gg\epsilon_d$. 
It is well known (see e.g. \cite{VZK}), that for this energy range the
solution of (\ref{eq1}) together with (\ref{bc1}) reads
\begin{equation}
\tanh(\theta(\bar{x})/4)=\tanh\left({i\pi\over8}\right)e^{k(\bar{x}-1)}
\end{equation}
where $k=d\sqrt{-2i\varepsilon/\cal{D}}$. By using obvious
substitutions and multiple-argument relations for hyperbolic
functions, we arrive at the following identity:
\begin{equation}
\label{help}
{\int_{\bar{x}}^1}d\bar{x}\,\sech(\theta_1(\bar{x}))=(1-\bar{x})-4\sqrt{\varepsilon_d\over\varepsilon}\int_0^{\Re(k)(1-\bar{x})}dy\,
q(y)/(1+q(y))^2
\end{equation}
where
$q(y)=4(3+2\sqrt{2}){e^{-2y}\sin^2y/\left(e^{-2y}+3+2\sqrt{2}\right)^2}$.

For calculating $D(\varepsilon)$ we can, as the integrand becomes
exponentially small for $y\ge\Re(k)\gg1$, take the upper bound to infinity,
such that it becomes a universal constant. From there we can
calculate the conductivity in this limit
\begin{equation}
\label{root}
\bar{\sigma}_N(T)=1+B\sqrt{\varepsilon_d\over T},
\end{equation}
where again $B=0.42$ is a universal constant.
The whole integral in the rhs of (\ref{help}) is very small compared to
one, so near the normal reservoir only the $1-\bar{x}$-term has to be
kept. As in this region $\nu$ is, apart from exponentially small
terms, constant, $\phi_0$ is linear in $\bar{x}$, so there $E_0$ is
constant. 

These results have a simple physical interpretation.
Superconductivity penetrates into the normal part up to
$\xi_N=\sqrt{D\over2\pi T}$, whereas the rest stays normal, so the total
voltage drops over a reduced distance $d-\xi_N$. Thus the resistance of
the structure is reduced according to the Ohm law. In terms of the
conductividy, this means 
\begin{equation}
\bar{\sigma}_N=1+B^\prime{\xi_N\over d}
\end{equation}
which is equivalent to (\ref{root}). 

For temperatures comparable to $\varepsilon_d$ the problem
can only be handled numerically. The results show excellent agreement
to the analytical calculations in the corresponding limits.

The numerical results (see fig 2) confirm, that for $\varepsilon_d\ll\Delta$ the
universal scaling with $T\over\varepsilon_d$ is excellently fulfilled,
as even the peak in the conductivity is systematically at
$T\approx5\varepsilon_d$ and the peak height of about 9\% seems to be
universal. This result also agrees with the recent numerical
analysis \cite{NazSt}. This peak becomes smaller if we take into
account the effect of finite $d_S$ keeping $d$ fixed (fig. 3)
The qualitative features, however, remain the same.

For the field distribution at $T\ll\varepsilon_d$ we can observe the
described nonmonotoneous behavior (see fig. 4). From $T\approx\varepsilon_d$ also a
minimum in the distribution shows up. For $T\gg\varepsilon_d$ the
electrical field is almost constant for $\bar{x}<1-\xi_N/d$, drops
rapidly for $\bar{x}>1-\xi_N/d$ and shows some interesting structure
around $\bar{x}=1-\xi_N/d$.

Perhaps the most interesting effect found here is the nonmonotoneous
dependence of the system conductance with temperature accompanied by a
nontrivial shape of the electric field penetrating in the N-layer. As $T$ is 
lowered proximity induced superconductivity in this layer becomes stronger
and stronger being more and more efficient in expelling the field
from the region adjacent to a superconductor. At very low temperatures this results in
a substantial suppression of this field in a big part of the sample. Inevitably
(the total potential drop is fixed!) this effect leads to an excess value
of the electric field far from the NS boundary as it is shown in fig.4. 
As it was already pointed out we can understand the effect of increase of 
the system conductance with increasing $T$ at very low temperatures as
a gap effect. At $T \sim \varepsilon_d$ the conductance reaches its maximum
and starts decreasing due to weakening of the proximity with further increase
of $T$.

Finally we would like to point out that the presence of tunnel barriers
at NN' interfaces entirely changes the conductance of the system.
If one lowers the barrier transparency the crossover
takes place to the behavior demonstrating monotonously decreasing
conductance with T (fig.5), which is typical for two serial NIS' 
tunnel junctions. Note that both types of
behavior, namely nonmonotoneous and monotonously decreasing with T
conductance have been observed in the experiment \cite{Petr}. Further
details will be presented elsewhere \cite{GWZ}.

We would like to thank G. Sch\"{o}n, C. Bruder and W. Belzig for useful 
discussions. This work was supported by
the Deutsche Forschungsgemeinschaft within the Sonderforschungsbereich 195.

\begin{figure}
\vspace{0.2cm}
\begin{minipage}[t]{15cm}
    \centerline{\psfig{figure=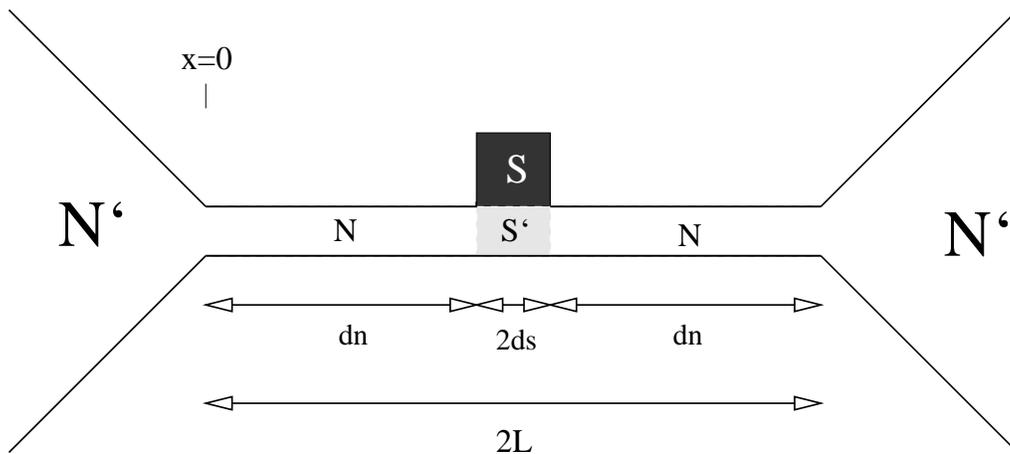,height=6cm}}
\end{minipage}
  \label{sketch}
      \caption{\footnotesize Physical setup. $N^\prime$ designates
normal reservoirs, $S$ superconductor, $S^\prime$ is effectly
superconducting due to the proximity to $S$, $N$ is the normal part
affected by the proximity to $S^\prime$}
\end{figure}

\begin{figure}
\vspace{0.2cm}
\begin{minipage}[t]{15cm}
    \centerline{\psfig{figure=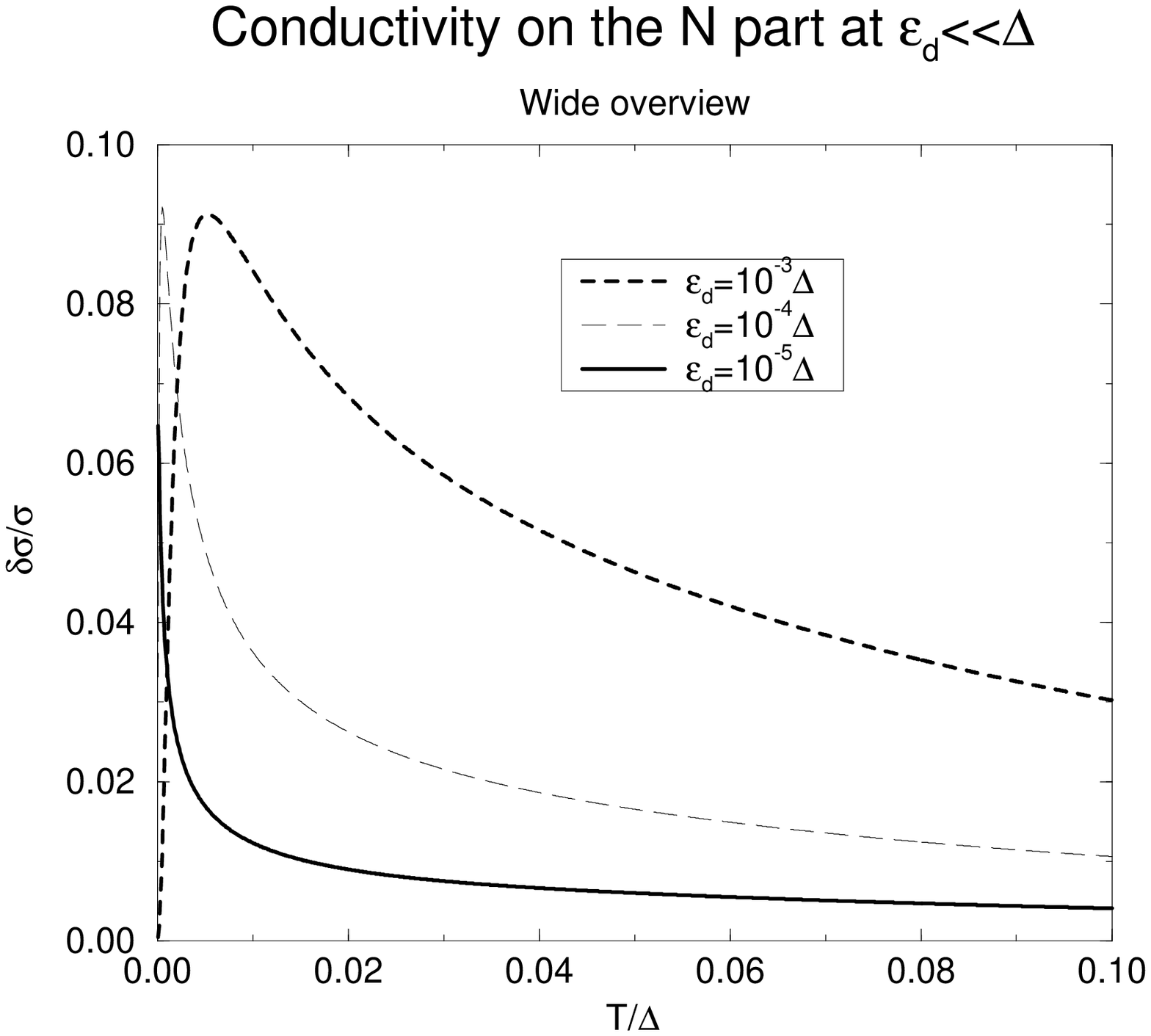,height=9cm}}
    \centerline{\psfig{figure=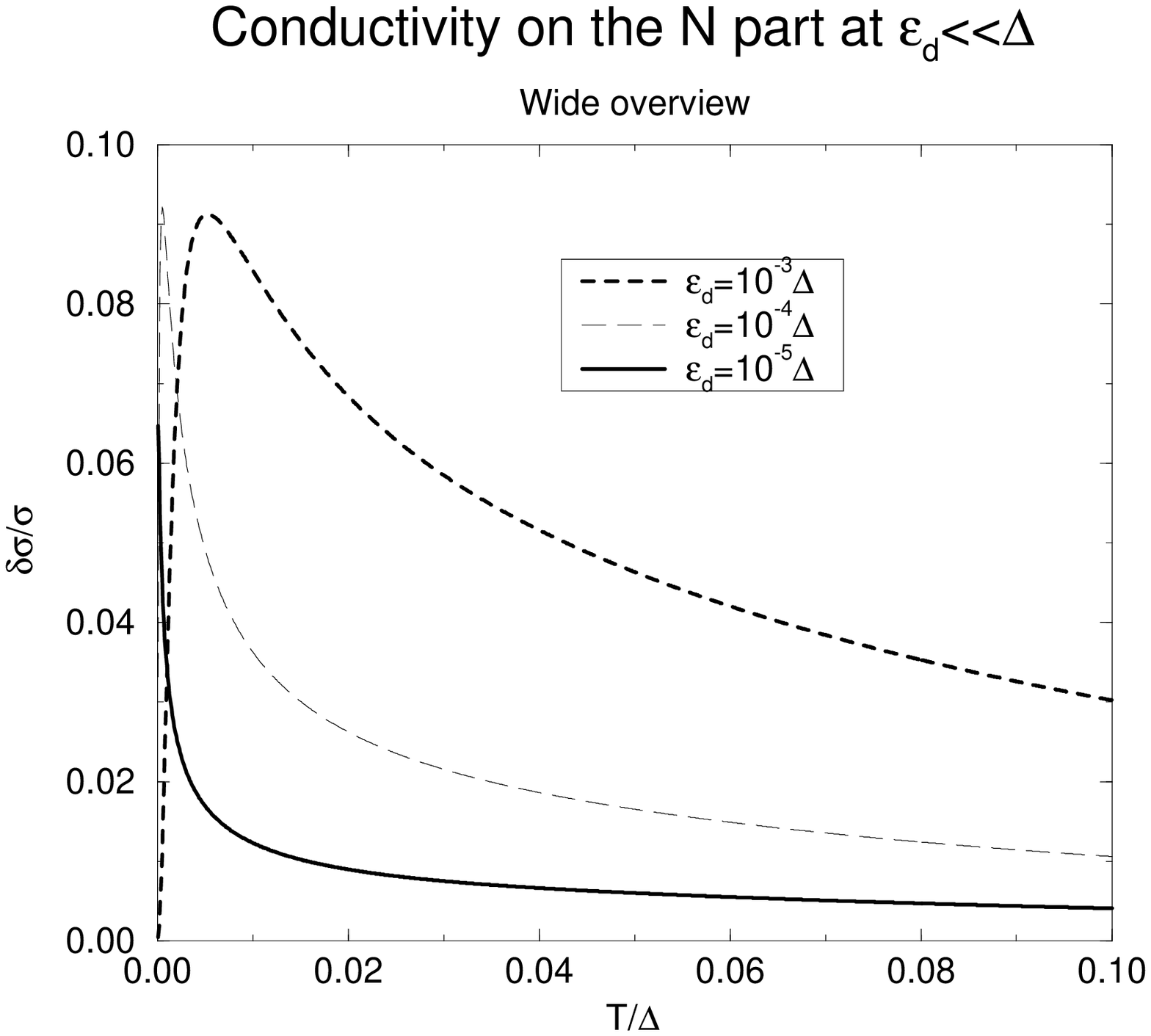,height=9cm}} 
\end{minipage}
  \label{cond}
      \caption{\footnotesize Temperature dependence of the
conductivity}
\end{figure}

\begin{figure}
\vspace{0.2cm}
\begin{minipage}[t]{15cm}
    \centerline{\psfig{figure=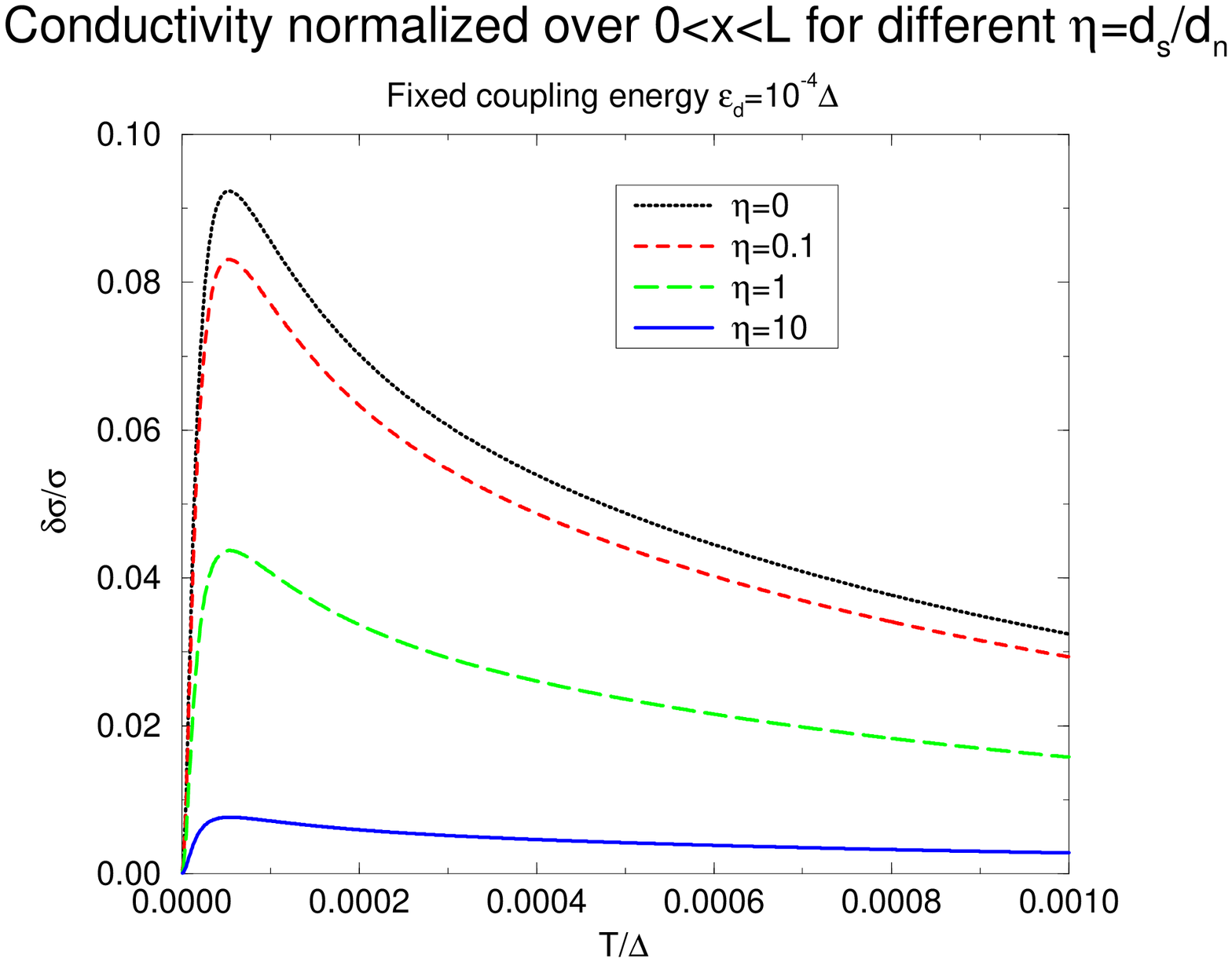,height=9cm}} 
\end{minipage}
      \label{eta}
      \caption{\footnotesize Influence of finite $d_s$}
\end{figure}

\vspace{0.6cm}
\begin{minipage}[t]{15cm}
    \centerline{\psfig{figure=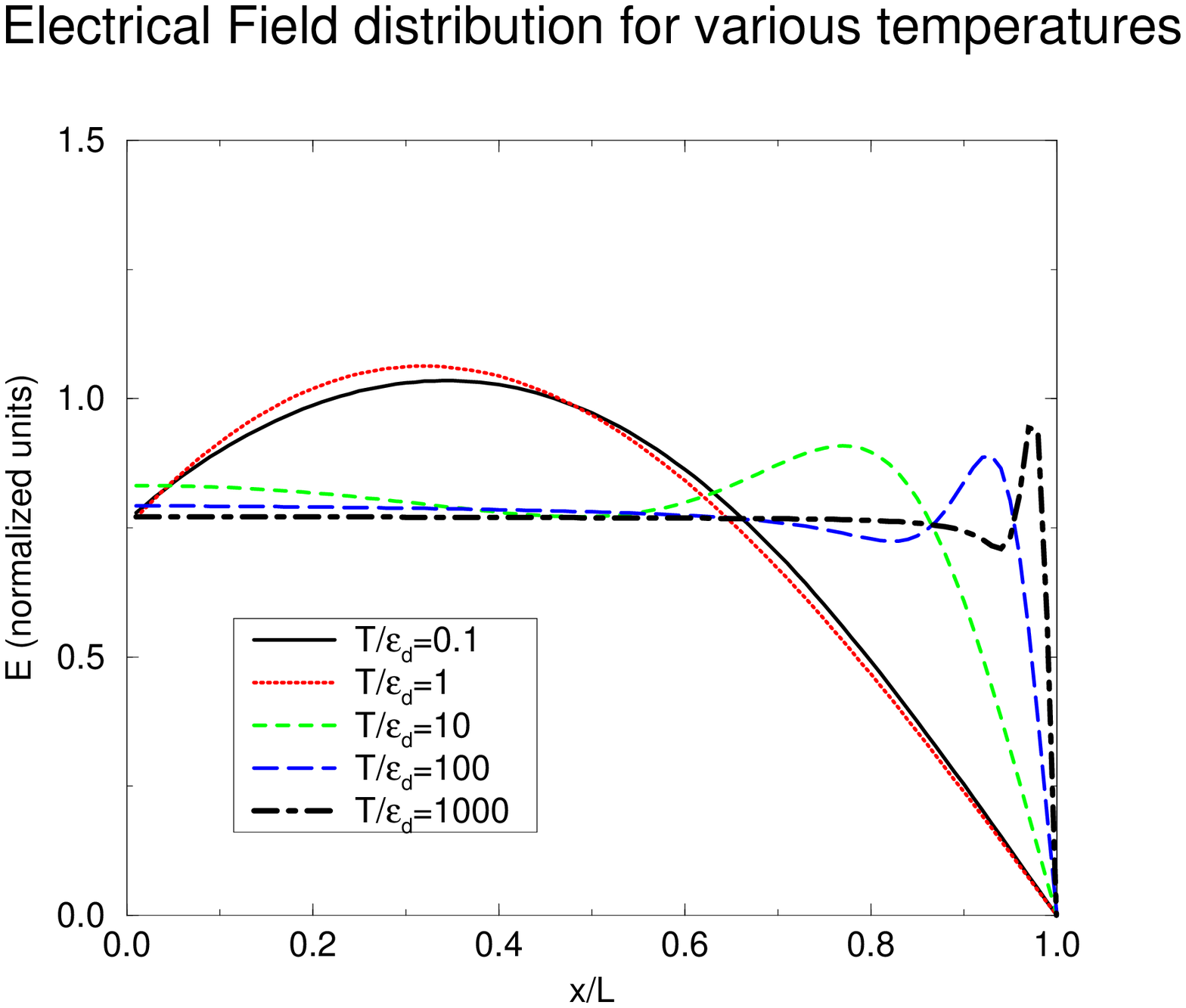,height=9cm}}
  \end{minipage}
  \hfill

\begin{figure}[htb]
      \label{efeld}
      \caption{\footnotesize Field distribution for various temperatures}
\end{figure}

\vspace{0.6cm}
\begin{minipage}[t]{15cm}
    \centerline{\psfig{figure=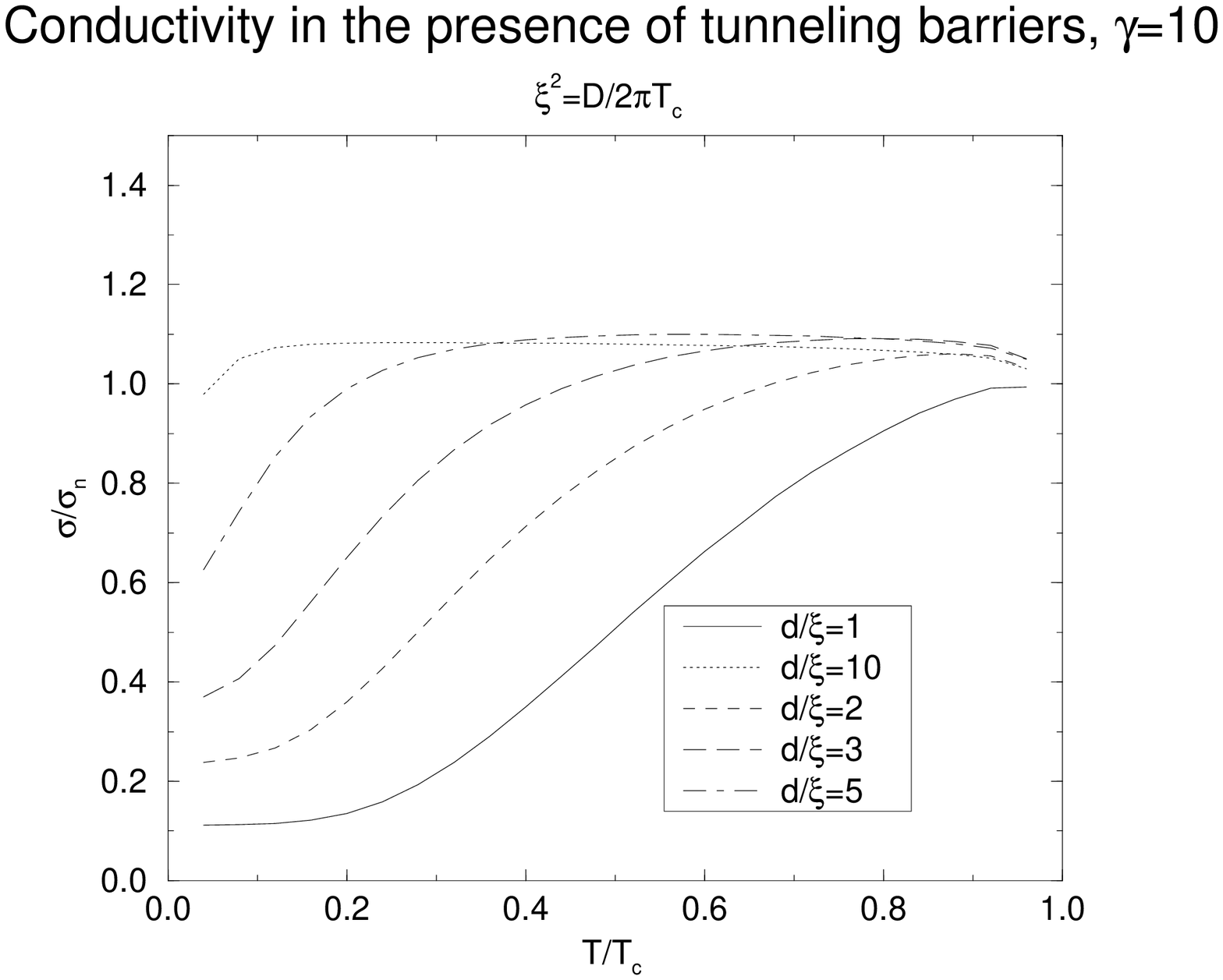,height=9cm}}
  \end{minipage}
  \hfill

\begin{figure}[htb]
      \caption{\footnotesize Destruction of the structure of $\sigma$ by the presence of strong tunneling barriers}
\end{figure}

\end{document}